\documentclass[twocolumn,showkeys,aps,nofootinbib,superscriptaddress]{revtex4}
\usepackage{graphicx}
\usepackage{adjustbox}
\usepackage{dcolumn}
\usepackage{bm}
\usepackage{multirow}
\usepackage{amsmath}
\usepackage[letterpaper, hmargin = {2cm}, vmargin = {2cm}]{geometry}
\usepackage{tablefootnote}

\usepackage{graphicx}
\usepackage{adjustbox}
\usepackage{dcolumn}
\usepackage{bm}
\usepackage{amsmath}
\usepackage{bbold}
\usepackage{multirow}
\usepackage{amsmath}
\usepackage[letterpaper, hmargin = {2cm}, vmargin = {2cm}]{geometry}
\usepackage{tablefootnote}

\usepackage[utf8]{inputenc}
\usepackage{graphicx} 
\usepackage{color}
\usepackage{amssymb}
\usepackage{amsmath}
\usepackage{amsfonts}
\usepackage{hyperref}
\usepackage{xcolor}
\usepackage{xfrac}

\usepackage{upgreek}
\usepackage{physics}

\def\A{\mathcal{A}}
\def\B{\mathcal{B}}

\usepackage[final]{changes}
\usepackage{mathrsfs}
\usepackage{float}

\begin{document}

\title{Structure of $^{43}$P and $^{42}$Si in a two-level shape-coexistence model}

\author{A.~O.~Macchiavelli}
\affiliation{Nuclear Science Division, Lawrence Berkeley National Laboratory, Berkeley, CA 94720, USA}

\author{H.~L.~Crawford }
\affiliation{Nuclear Science Division, Lawrence Berkeley National Laboratory, Berkeley, CA 94720, USA}

\author{ C.~M.~Campbell}
\affiliation{Nuclear Science Division, Lawrence Berkeley National Laboratory, Berkeley, CA 94720, USA}

\author{ R.~M.~Clark }
\affiliation{Nuclear Science Division, Lawrence Berkeley National Laboratory, Berkeley, CA 94720, USA}

\author{ M.~Cromaz }
\affiliation{Nuclear Science Division, Lawrence Berkeley National Laboratory, Berkeley, CA 94720, USA}

\author{ P.~Fallon }
\affiliation{Nuclear Science Division, Lawrence Berkeley National Laboratory, Berkeley, CA 94720, USA}

\author{ I.~Y.~Lee }
\affiliation{Nuclear Science Division, Lawrence Berkeley National Laboratory, Berkeley, CA 94720, USA}

\author{A. Gade}
\affiliation{National Superconducting Cyclotron Laboratory, East Lansing, Michigan 48824, USA}

\author{A.~Poves} 
\affiliation{Departamento de F\'isica Te\'orica  and IFT-UAM/CSIC, \mbox{Universidad
 Aut\'onoma de Madrid, 28049 Madrid, Spain}}
 
\author{E.~Rice}
 \affiliation{Department of Physics and Astronomy, Ohio University, Athens, OH}
\affiliation{Nuclear Science Division, Lawrence Berkeley National Laboratory, Berkeley, CA 94720, USA}

\date{\today}
 
\begin{abstract}
Exclusive cross sections for the $^{43}$P$(-1p)^{42}$Si reaction to the lowest $0^+$ and $2^+$ states, measured at NSCL with GRETINA and the S800, are interpreted in terms of a two-level mixing (collective) model of oblate and prolate co-existing shapes. Using the formalism developed for deformed nuclei we calculate the spectroscopic amplitudes and exclusive cross-sections in the strong coupling limit, where for $^{43}$P the schematic
wavefunction includes the coupling of the Nilsson [211]$\frac{1}{2}$ proton orbit. Good agreement with the experimental data is obtained when the amplitude of the oblate configuration is $\gtrsim$ 80\%, suggesting that both nuclei are predominantly oblate, in line with theoretical expectations. 
\end{abstract}


\maketitle

\section{Introduction}

Moving away from the valley of beta stability, the neutron-proton imbalance results in a large asymmetry energy which is ultimately   responsible for the neutron dripline, beyond which isotopes  with additional neutrons are not bound.  For the bound systems, the same nucleon imbalance, affects the details of the residual nuclear interaction, leading to altered ground and excited state properties for exotic nuclei as compared to those observed in stable isotopes. Strong evidence can be found in the literature highlighting the major breakdowns of the spherical shell-model with respect to where shells appear in 20$<N<$30 and 10$<Z<$20 nuclei~\cite{Sor2008,Ot20}.  Exploration of these changes is fundamental to our understanding of nuclear structure, is critical for predicting the location of the neutron dripline and has a potentially significant impact on r-process nucleosynthesis.  

The neutron-rich $N$=28 region is ideal (and the heaviest region experimentally accessible at the neutron drip-line) for exploring such structural changes. The $N$=28 nuclei below $Z$=20 show examples of altered single-particle structure, collective degrees of freedom, and potential impacts of weak binding on structure as the dripline is reached near $^{40}$Mg.  

 Whereas the gap between the 0f$_{7/2}$ and the 1p$_{3/2}$ orbits in $^{41}$Ca results from the combination of the spin-orbit coupling and the $\ell^2$ term of the spherical mean field, and amounts to $\approx$ 2 MeV, it evolves to $\approx$4.5~MeV in $^{48}$Ca due to the fact that the T=1 monopole interaction of the 0f$_{7/2}$ neutrons among themselves is more attractive than the interaction  between the the 0f$_{7/2}$ and the 1p$_{3/2}$ neutrons. It is this increased gap that makes $^{48}$Ca a doubly-magic nucleus. 
 Below $^{48}$Ca, as protons are removed from the $d_{3/2}$ orbital, the effects of the $\Delta \ell $=1 spin-flip interaction effectively reduces the $\nu f_{7/2}$-$\nu p_{3/2}$ gap. The spacing between the proton $\pi d_{5/2}$, $\pi s_{1/2}$ and $\pi d_{3/2}$ orbitals are also narrowed at $N$=28~\cite{ Gad2006,Gau06, Ril2008,  Sor2008, Now09,Uts2012,Ot20}.  With this, the $N$=28 isotones from  Ar to Mg display the effects of a varying $N$=28 gap, with development of deformation and changes in shape~\cite{Heyde1}.  

 An understanding of the evolution of shapes along $N$=28 has been proposed, in which the quadrupole force  mixing between $m$ substates in the nearly-degenerate proton orbitals, and the resulting change in occupancy of the deformation-driving $\ell$=2, $m$ = $\pm\frac{1}{2}$ substates, results in a tensor force driven Jahn-Teller effect~\cite{Uts2012}, if seen in the intrinsic frame, or simply a manifestation of Elliott's SU(3) symmetry in the laboratory frame. This mixing of the proton $\pi d_{5/2}$ and $\pi s_{1/2}$ levels suggests changing shapes, prolate-oblate-prolate, in $^{44}$S, $^{42}$Si, and $^{40}$Mg respectively. While the available experimental data appear consistent with this scenario, a 'direct' confirmation of the dominant shapes and the potential shape coexistence in the region of $N \sim 28$ remains to be studied in more detail.
 
 In fact the recent study of the one proton knockout reaction from $^{43}$P to $^{42}$Si~\cite{Gad20} has shown that large scale shell model calculations with the SDPF-U~\cite{Now09} and SDPF-MU~\cite{Uts2012} effective interactions, that have been successfully used in this region,  both  predict an oblate ground state band in agreement with the experiment, while they differ in the location of the
 coexisting prolate band which is better reproduced by the SDPF-MU interaction ~\cite{Gad20}. 
 
While the shell-model has been applied broadly and successfully in this region, what has not been fully explored is the success, or lack thereof, of the arguably more intuitive Nilsson model.  We consider here a description of the $^{43}$P$(-1p)^{42}$Si reaction in the framework developed for deformed nuclei and explore the validity of this description.

\section{The two-level model}

The available experimental data in comparison with shell model calculations indicate the low-energy structures of $^{44}$S, $^{42}$Si, and $^{40}$Mg are each likely dominated by two major co-existing configurations,  spherical and prolate in the case of $^{44}$S, and  oblate and prolate  for both $^{42}$Si, and $^{40}$Mg. This suggests that a two-state (shape) mixing model can provide a useful description of their structure. For example, Ref.~\cite{For10} used such a model to describe the low-energy structure of $^{44}$S. From our previous application of this model in the description of the two-proton knockout reactions, $^{44}$S(-2p)$^{42}$Si and $^{42}$Si(-2p)$^{40}$Mg, a picture  
emerged in which the dominant nuclear shape changes from prolate in $^{44}$S, to oblate in $^{42}$Si, and returning to prolate in $^{40}$Mg~\cite{Cra14}, in line with the qualitative arguments.  While the first $\gamma$-ray spectroscopic study of $^{40}$Mg~\cite{Cra19} may point to intriguing effects due to the  weak binding of the  two valence neutrons, its deformation appears consistent with the prolate trend along the Mg isotopic chain up to $^{38}$Mg~\cite{Doo13,Tsu20}. 

Here, we follow Ref.~\cite{Cra14} and take advantage of the continued experimental progress in this region, and apply a two-state mixing model to interpret the results of Ref.~\cite{Gad20}.  We consider specifically the cross-sections from the 1/2$^{+}$ ground state of $^{43}$P into the 0$^{+}_{1}$ ground state of $^{42}$Si, the 2$^+_{1}$ state at 737(8)~keV, the tentatively assigned (0$^{+}_{2}$) state at 2150(13)~keV and a tentative (2$^{+}_{2}$) state at $\approx$ 3 MeV.  In $^{43}$P and $^{42}$Si, guided by shell model calculations, we expect the co-existence of oblate ($O$) and prolate ($P$) deformed components to be dominant in the low-energy configurations. 

Consider that the ground state wave-function of $^{42}$Si is described in the simple form:
\begin{align}
\ket{0_1^+} =  \alpha\ket{O}+\beta\ket{P}
\label{eq:42SiWaveGS}
\end{align}
 \noindent
with the corresponding orthogonal $0^+_2$ state  
\begin{align}
\ket{0_2^+} =  -\beta\ket{O}+\alpha\ket{P}
\label{eq:42SiWaveSecond0}
\end{align}
 \noindent
 Similarly,  we have for the lowest 2$^+$ states:
 \begin{align}
\ket{2_1^+} =  \alpha'\ket{ O}+\beta'\ket{P}
\label{eq:42SiWaveFirst2}
\end{align}
 \noindent
\begin{align}
\ket{2_2^+} =  -\beta'\ket{O}+\alpha'\ket{P}
\label{eq:42SiWaveSecond2}
\end{align}
 \noindent
 
\begin{figure}[h]
\begin{centering}
\includegraphics[trim=0 30 40 70, clip,width=0.9\columnwidth, angle=90]{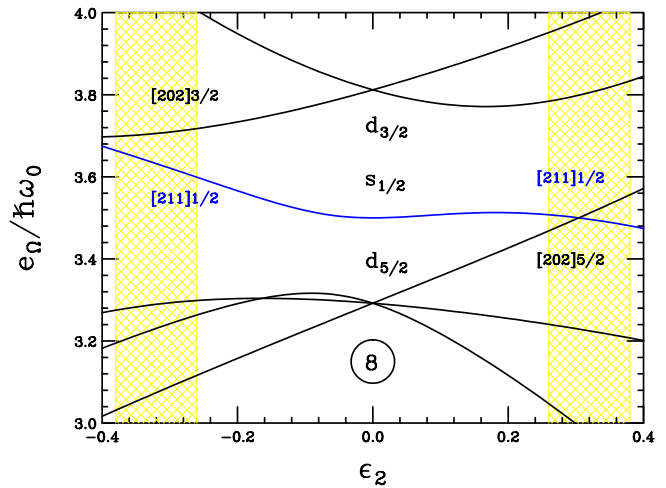}
\caption{ Proton Nilsson orbitals originating from the $sd$ shell and relevant for the structure of $N=28$ nuclei.  Specifically for the case of  $^{43}$P, the [211]1/2 level is highlighted in blue and the shaded yellow area indicates the anticipated regions of deformation (See text).}.  
\label{fig:nilsson}
\end{centering}
\end{figure}

In $^{43}$P, an inspection of the Nilsson diagram~\cite{Sven,Rag} relevant for $Z$=15, as is shown in Fig.~\ref{fig:nilsson}, suggests that the $\ket{1/2^+}$ ground state could be associated to the [211]$\frac{1}{2}$ at both prolate and oblate deformations.
More generally, in the two-level model we have:
\begin{align}
\ket{1/2_1^+} = \A \ket{[211]\frac{1}{2}} \otimes \ket{O}  + \B \ket{[211]\frac{1}{2}} \otimes  \ket{P}
\label{eq:35Pgs}
\end{align}
where it is implicitly understood that the [211]$\frac{1}{2}$ level wavefunctions are those at the appropriate ($O$ or $P$) deformations. We indicated with the shaded yellow areas a range of deformations that one may anticipate based on the 2$^+_1$ energies in $^{42}$Si, 787 keV, and $^{44}$S, 1329 keV, using the Migdal formula for the moment of inertia~\cite{Migdal}. 
In the spherical $|j,\ell\rangle$ basis the [211]$\frac{1}{2}$ level wavefunction takes the form:

\begin{align}
|[211]\tfrac{1}{2}\rangle=  C_{1/2,0} |s_{1/2}\rangle+ C_{3/2,2} |d_{3/2}\rangle+  C_{5/2,2}|d_{5/2}\rangle
\label{eq:nilsson211}
\end{align}
 where the Nilsson coefficients $C_{1/2,0}, C_{3/2,2}$ and $C_{5/2,2}$ are functions of the deformation parameter $\epsilon_{2}$.  The Nilsson coefficients were calculated, taking $\mu$=0 and $\kappa$= 0.105~\cite{Sheline}.

\section{The $^{43}$P(-1p)$^{42}$Si reaction}
\begin{figure}[h]
\begin{centering}
\includegraphics[trim=0 30 40 70, clip,width=0.9\columnwidth, angle=90]{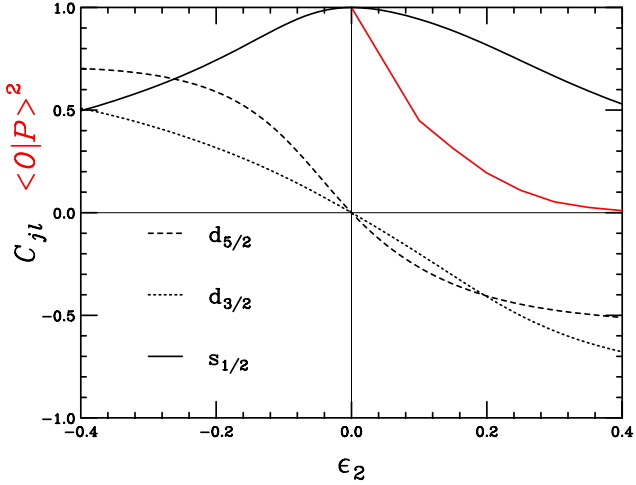}
\caption{Nilsson amplitudes for the [211]$\frac{1}{2}$ level as a function of deformation. The red line shows the estimated core overlap $\langle O|P \rangle$. } 
\label{fig:nilssonAmplitudes}
\end{centering}
\end{figure}

We apply the formalism reviewed in Ref.~\cite{Elbek} to the proton knockout reaction.  In the strong coupling limit of the Particle Rotor Model (PRM)~\cite{Rag}, the spectroscopic factors ($S_{i,f}$) from an initial ground state $|I_iK_i\rangle$ to a final state $| I_f K_f\rangle$ (with $K_f = 0$) can be written in terms of the Nilsson amplitudes:

\begin{equation} 
\begin{split}
\theta_{i, f  }(j\ell,K)& = \sqrt{2}\langle I_{i}j  K \Omega_\pi  | I_{f} 0\rangle C_{j,\ell} \langle\phi_f|\phi_i\rangle \\
S_{i, f  }& =  V^2_{\Omega_\pi}\theta_{i, f  }^2(j\ell,K)
\end{split}
\label{eq:specFactor}
\end{equation}
where $\langle I_{i}j  K \Omega_\pi  | I_{f} 0\rangle$ is a Clebsch-Gordan coefficient, $C_{j,\ell}$ is the Nilsson wavefunction amplitude, $\langle\phi_f|\phi_i\rangle$ is the core overlap between the initial and final states, and $ V_{\Omega_\pi}$ the occupancy of the Nilsson level. For this case we assume for the overlaps: $\langle O_f|O_i \rangle$= $\langle P_f|P_i \rangle$=1 and  $\langle O_f|P_i \rangle$=$\langle P_f|O_i \rangle$=0.  An estimate~\cite{TNA1, TNA2} of $\langle O |P \rangle $\footnote{Here, we consider that both configurations have the same value of $|\epsilon_2|$.} vs. $\epsilon_2$ in the range of interest, shown in Fig.~\ref{fig:nilssonAmplitudes}, justifies the assumption above. As can also be seen, the amplitudes $C_{j,\ell}$ evolve very smoothly; for the analysis to follow the values of $C_{j,\ell}$ are parameterized by fifth-order polynomial functions for ease of the calculations.

Following the proton knockout into $^{42}$Si,  the $0^+$ states can only be reached through the $C_{1/2,0}$ component of the Nilsson wavefunction in Eq.~\ref{eq:nilsson211}, while the $2^+$ states are populated through both $C_{3/2,2}$ and $C_{5/2,2}$ components.  Taking into account the expressions in Eqs. 1-7,  we obtain the following spectroscopic amplitudes for the $0^+_1$ and $0^+_2$ levels:
\begin{equation} 
\begin{split}
 \theta_{1/2, 0^+_1} (0,1/2)& =  \alpha \A C_{1/2,0}(O) + \beta \B C_{1/2,0}(P)\\
 \theta_{1/2,0^+_2 }(0,1/2)& =  -\beta \A C_{1/2,0}(O) + \alpha \B C_{1/2,0}(P)
\end{split}
\label{eq:eq4}
\end{equation}
from which we calculate the cross-sections~\cite{Tosab}:
\begin{equation} 
\begin{split}
 \sigma_{th}(0_{1(2)}^+)& =  R \left(\frac{A}{A-1}\right)^2\theta^2_{1/2, 0^+_{1(2)}} (0,1/2) \sigma_{sp}^{0,1/2, L(H)}\\
\end{split}
\label{eq:eq4}
\end{equation}
 Similarly, we obtain for $2^+_1$ and $2^+_2$ levels:
\begin{equation} 
\begin{split}
 \theta_{1/2,2^+_1 }(2,3/2)& =  \alpha' \A C_{3/2,2}(O) + \beta' \B C_{3/2,2}(P)\\
 \theta_{1/2,2^+_2 }(2,3/2)& =  -\beta' \A C_{3/2,2}(O) + \alpha' \B C_{3/2,2}(P)\\
\end{split}
\label{eq:eq4}
\end{equation}
\begin{equation} 
\begin{split}
 \theta_{1/2,2^+_1 }(2,5/2)& =  \alpha' \A C_{5/2,2}(O) + \beta' \B C_{5/2,2}(P)\\
 \theta_{1/2,2^+_2 }(2,5/2)& =  -\beta' \A C_{5/2,2}(O) + \alpha' \B C_{5/2,2}(P)\\
\end{split}
\label{eq:eq4}
\end{equation}
and the cross-sections:
\begin{equation} 
\begin{split}
 \sigma_{th}(2_{1(2)}^+)& =  R  \left( \frac{A}{A-1} \right)^2 \Big[ \theta^2_{1/2, 2^+_{1(2)}} (2,3/2) \sigma_{sp}^{2,3/2, L(H)} \\
 & +    \theta^2_{1/2, 2^+_{1(2)}} (2,5/2)\sigma_{sp}^{2,3/2, L(H)}\Big]\\
\end{split}
\label{eq:eq4}
\end{equation}
In Eqs. (9) and (12), $R$ is a quenching factor which includes the occupancy $V^2_{1/2}$, and $\sigma_{sp}^{\ell, j, L(H)}$ are the single-particle cross-section associated with a given orbital and final state binding energy, denoted here by $L$ for the lower energy final state (0$_{1}^{+}$ and 2$_{1}^{+}$) and $H$ for the higher energy final states (0$_{2}^{+}$ and 2$_{2}^{+}$).  The single-particle cross-sections adopted here are those in Ref.~\cite{Gad20}, which were calculated following the one-nucleon removal methodology outlined in Ref.~\cite{Gad08}.

\section{Results}

\begin{figure}[h]
\begin{centering}
\includegraphics[width=7cm, angle= 90]{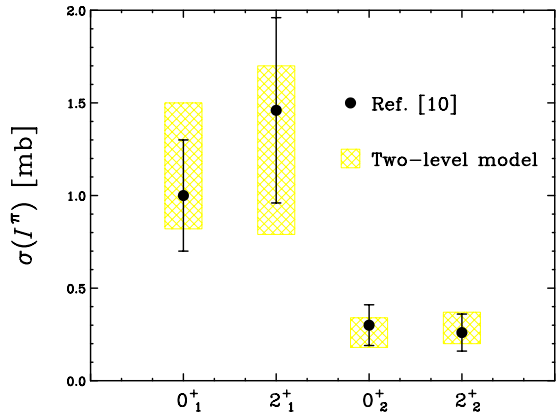}
\caption{Results of the calculation (yellow bars) to reproduce the experimental cross-sections (black data points).  The reproduction is very good, with the asymmetric error bars for the calculation coming as a result of the constraint $\A$,$\alpha\in[0,1]$. } 
\label{fig:fit}
\end{centering}
\end{figure}

Having discussed the ingredients entering in the calculations, we proceed to compare our predictions with the data of Ref.~\cite{Gad20}. To limit the number of parameters, we make the simplifying and reasonable assumption that, $\alpha=\alpha'$ ($\beta=\beta'$) which leaves us to fit $\A$, $\alpha$, $R$ and the absolute value of the deformation $\epsilon_{2}$, on which the Nilsson coefficients depend.  With these assumptions, we are not constraining the deformations of $^{43}$P and $^{42}$Si to be the same, and the system of equations is not under-constrained.
With four parameters and four data points, we reproduce very closely the experimental cross-sections under consideration, as seen in Figure~\ref{fig:fit}. Mathematically we would expect to reproduce exactly the experimental data - the small discrepancy and asymmetric error bars in our calculation are representative of the physical constraint of $\A$,$\alpha\in[0,1]$.  However, the primary question is whether the solutions obtained correspond to physically reasonable parameters.  We find $R$ = 0.33(7) and $|\epsilon_{2}| = 0.26(4)$. The magnitude of deformation, $\epsilon_{2}$, is well in line with expectations based on the neighbouring isotopes.

With respect to the values of $\A$ and $\alpha=\alpha'$, we find three solutions, or minima in the $\chi^2$ surface under the constraints that $\A$ and $\alpha$ are bound within the limits of [0,1].  Two solutions correspond to $(\A, \alpha) = (0^{+0.04}_{-0}, 0.45(7)$ and $(0.45(7), 0^{+0.04}_{-0})$, while the deepest minimum is located at $(\A,\alpha)=(0.90(3), 1^{+0}_{-0.001})$.  In other words, the experimental cross-sections are reproduced for the scenario in which both $^{43}$P and $^{42}$Si are predominantly oblate, or a scenario in which one of the two systems is highly mixed, while the other is essentially completely prolate.  Unfortunately, with only four data points at this time, it is not possible to further constrain the model fit.  However, we can conclude, taking into account the predictions of the shell-model which suggest that $^{42}$Si is predominantly oblate, that $^{43}$P is likely similarly deformed.

As a final point, it is interesting to note that, if we consider our quenching factor $R$ as akin to the reduction factor $R_{s}$ observed in knockout reactions~\cite{Tos14,Tos21} (which are referenced to 
large-scale shell-model calculations), we find good agreement with what would naively be expected from the established trend of Ref.~\cite{Tos21} ($R_s$=0.36) and the value of 0.33(2) reported in Ref.~\cite{Gad20}.  Going further, a BCS calculation of all the Nilsson states in Fig.~\ref{fig:nilsson}, gives an occupancy $V^2_{1/2} \approx 0.5$ for the [211]$\frac{1}{2}$ level. Having considered this as a part of our quenching factor $R$, we can deduce an effective suppression of $\approx 0.68$. Thus, in comparison to the shell model, it appears that the Nilsson wavefunction perhaps more completely captures the fragmentation of the spherical single-particle strength due to the quadrupole force in these deformed systems.
\added{With a BCS proton gap, $2\Delta \approx 3$ MeV, the additional strength observed above 3 MeV can be associated with fragments of the $d_{5/2}$ orbit present in two quasi-particle states in $^{42}$Si  populated by proton knockout from Nilsson levels below the Fermi surface ($\lambda/\hbar\omega_0 \approx$ 3.5 in Fig.~\ref{fig:nilsson}) in $^{43}$P.}
\section{Conclusion}

We have interpreted the results of a recent study of the $^{43}$P$(-1p)^{42}$Si reaction~\cite{Gad20} in terms of a two-level mixing model of oblate and prolate co-existing shapes. For $^{43}$P the schematic
wavefunction includes the coupling of the Nilsson [211]1/2 proton orbit and using the formalism developed for deformed nuclei~\cite{Elbek} we calculate the spectroscopic amplitudes and exclusive cross-sections to the $0^+_{1,2}$ and $2^+_{1,2}$ states. A realistic solution that reproduces the experimental data is obtained, which suggests that both $^{42}$Si and $^{43}$P are dominated by oblate configurations in their ground states. The quadrupole invariants method~\cite{Pov20} applied to the SDPF-U~\cite{Now09} interaction $^{42}$Si calculations predicts, $\epsilon_2=0.32(7)$ and $\gamma= 47(13)^\circ$ for the ground state of the lower band and $\epsilon_2=0.34(6)$ and $\gamma= 15(15)^\circ$ for the higher band, consistent  with our Nilsson results.
Finally, although  strong coupling   appears to capture the underlying nuclear structure, perhaps a further refinement on the basis of a full PRM calculation~\cite{Semmes} might be considered.  \\

\begin{acknowledgments}
 This material is based upon work supported by the U.S. Department of Energy, Office of Science, Office of Nuclear Physics under Contracts No. DE-AC02-05CH11231 (LBNL), DE-SC0020451 (NSCL). AP work is supported in part by the Ministerio de Ciencia, Innovaci\'on y Universidades (Spain), Severo Ochoa Programme SEV-2016-0597 and grant PGC-2018-94583. \\
 \end{acknowledgments}


\end{document}